\documentclass[12pt]{article}
\usepackage{jheppub}
\usepackage{amsfonts,ulem,bbold}
\usepackage{amsmath,amssymb, braket}
\usepackage{subcaption}

\usepackage{graphicx}
\usepackage{caption, threeparttable}

\newcommand{\beqn}{\begin{eqnarray}}
\newcommand{\eeqn}{\end{eqnarray}}
\newcommand{\be}{\begin{equation}}
\newcommand{\ee}{\end{equation}}

\title{Black Hole Lattices Under the Microscope}   
\author{Ingemar~Bengtsson,}
\author{Irina~Galstyan}
\affiliation{Department of Physics \& 
        The Oskar Klein Centre,\\
        Stockholm University, AlbaNova University Centre, 
        SE-106 91 Stockholm, Sweden}
\emailAdd{ibeng@fysik.su.se}
\emailAdd{irina.galstyan@fysik.su.se}

\abstract{It is known how to choose initial data for Einstein's equations 
describing an arbitrary number of black holes at a moment of time symmetry. 
This idea has been used to give insight into the cosmological averaging 
problem. We study the local curvature of the initial data space, for 
configurations of 8, 120, or 600 black holes obtained by choosing points 
either regularly or randomly on a 3-sphere. We conclude that the asymptotic 
regions are remarkably close to that of Schwarzschild, while the region 
in between shows interesting behaviour. The cosmological back reaction as 
defined in the recent literature is actually a bit smaller for the random 
configurations.}

\toccontinuoustrue

\begin{document} 
\maketitle
\flushbottom
\section{Introduction}
\label{intro}

\noindent Before the First Golden Age (of General Relativity) Charles Misner 
proposed that the study of time-symmetric initial data should be used to provide 
intuition about Einstein's equations, much in the same way as the study of 
electrostatics prepares the student for understanding Maxwell's equations. 
Indeed, Misner \cite{Misner}, and Brill and Lindquist \cite{Brill}, soon 
demonstrated the soundness of this idea. It has experienced several revivals, 
notably in the study of the Penrose inequality \cite{Gibbons}, in the 
early stages of numerical relativity \cite{DeWitt}, and more 
recently in order to produce thought provoking toy models of the cosmological 
averaging problem \cite{Kjell, BandK}. These toy models consist of initial data 
for a large number of black holes distributed on a space that in a sense 
approximates a round 3-sphere, and go under the name of ``black hole lattices''. 
They differ from the black hole lattices of Lindquist and Wheeler \cite{Wheeler} 
in that they rely entirely on exact solutions of Einstein's vacuum equations. 

In this paper we try to understand what the black hole lattices really look 
like, and we study how their local curvature behaves. It should be kept in mind 
that a surface can provide a very good approximation of a round sphere in 
one sense, even though it looks very different from a round sphere under the 
microscope---a point made very nicely by Green and Wald \cite{Green} in connection 
with the cosmological averaging problem. We do not discuss the dynamical aspects 
of the problem at all, and just remark that this has been done elsewhere, 
with interesting results \cite{BandK, Daniele, khb, Daniele2}. 

The starting point of Misner's ``geometrostatics'' is the observation that, in 
the absence of matter, time-symmetric initial data $(g,K)$ for Einstein's 
equations are obtained by solving the equations  

\begin{equation} R(g) = 0 \ , \hspace{8mm} K_{ij} = 0 \ . \end{equation}

\noindent This can be simplified by insisting that the 3-metric $g$ be 
conformal to flat space, or equivalently conformal to the 3-sphere. The Schwarzschild 
solution admits initial data of this kind, on a space which is topologically 
a twice punctured 3-sphere. The two punctures correspond geometrically to the 
two asymptotic regions. Brill and Lindquist \cite{Brill} made a detailed study of 
the case where space has an arbitrary number $N$ of punctures, giving rise to a 
solution with $N$ asymptotic regions. Alternatively, provided that the $N > 2$ 
punctures are distributed in a reasonable way, this can be looked at from 
the inside as a solution describing a universe containing $N$ black holes. 
There exists a precise theorem, due to Korzy\'nski \cite{Korzynski}, which 
says that the resulting metric will approximate the round metric on the 
3-sphere increasingly well, over an increasing fraction of the 3-sphere, in 
the limit of large $N$. Our main purpose in this paper is to study the 
behaviour of the second derivatives of the metric. We do this by studying 
how the local curvature behaves in some examples with a reasonably large 
number of black holes. 

In section \ref{BL} of this paper we describe the geometrostatical initial data as simply 
as we can, and provide explicit formulas for the Ricci tensor and for the curvature 
scalar of a two dimensional cross-section while keeping the number and the positions 
of the black holes arbitrary. In section \ref{Platonic} we describe some special configurations 
defined by four dimensional Platonic bodies. We also specify a procedure for 
choosing the positions of the black holes `at random', and study the distribution 
of ADM masses that results. In section \ref{lattice} we draw some exact pictures of the resulting spaces, following Clifton et 
al. \cite{Kjell} but with a slight twist. In section \ref{local} we study the behaviour of 
the local curvature in various configurations. Section \ref{concl} gives our concluding remarks. 

For readers who want a thorough understanding of the subject and how it fits 
into an attempt to understand the cosmological averaging problem, we are 
in the fortunate position of being able to refer to an excellent and up to 
date review \cite{revy}.


\section{ Brill--Lindquist data and the 3-sphere}
\label{BL}
\noindent At a moment of time-symmetry the extrinsic curvature of an initial data 
slice vanishes, and Einstein's constraint equations reduce to the statement that the 
3-metric should have a vanishing curvature scalar. Misner proposed 
a further restriction to conformally flat spaces, so that the physical metric 
$g$ is given by 

\begin{equation} g_{ab} = \omega^4\hat{g}_{ab} \ , \end{equation}

\noindent where the metric $\hat{g}$ is taken to be either flat or to be 
that of the round 3-sphere. From well known formulas \cite{Wald} it follows 
(in dimension 3) that 

\begin{equation} R = -\frac{8}{\omega^5}\tilde{\Delta}\omega \ , \end{equation}

\noindent where the conformally invariant Laplace operator 

\begin{equation} \tilde{\Delta} = \hat{g}^{ab}\hat{\nabla}_a\hat{\nabla}_b 
- \frac{1}{8}\hat{R} \end{equation}

\noindent appears on the right hand side. For the unit 3-sphere $\hat{R} = 6$. 

We will describe the round 3-sphere using dimensionless embedding coordinates $X^a = 
(X,Y,Z,U)$, so that 

\begin{equation} ds^2_{\rm sphere} = \frac{m^2}{4}(dX^2 + dY^2+ dZ^2 + dU^2) \ , 
\hspace{8mm} X^2+Y^2+Z^2+U^2 = 1 \ . \label{skala} \end{equation}

\noindent We set the dimensionful constant $m = 2$, and stick to this until we have 
to select a scale for the black hole lattices in the next section. In this way we 
make maximum use of the fact that Einstein's vacuum equations define a scale 
invariant theory.  

The four components of 
$X^a$ can be parametrized by (dimensionless) stereographic or geodesic polar 
coordinates as  

\begin{eqnarray} X = \frac{2x}{1+r^2}= \cos{\phi}\sin{\theta}\sin{\chi}  \hspace{8mm} 
Y = \frac{2y}{1+r^2} = \sin{\phi}\sin{\theta}\sin{\chi}  \nonumber \\ \label{embcoord} \\ 
Z = \frac{2z}{1+r^2}= \cos{\theta}\sin{\chi}  \hspace{10mm} 
U = \frac{1-r^2}{1+r^2} = \sin{\phi}\sin{\theta}\sin{\chi} \ .  \nonumber \end{eqnarray} 

\noindent We try to avoid intrinsic coordinates as far as possible.

In order to find a non-trivial solution of the conformal Laplace equation we 
puncture the 3-sphere at $N$ locations, and find the solution \cite{Misner, Brill} 

\begin{equation} \omega = \sum_{i=1}^N\omega_i \ , \hspace{8mm} \omega_i 
= \frac{1}{\sqrt{1 + X\cdot {\bf X}_i}} \ . \label{sol} \end{equation}

\noindent Here ${\bf X}_i^a$ is a constant unit 4-vector. One can introduce different 
coefficients in front of the $N$ terms, but for simplicity and definiteness we set 
all of these integration constants equal.  

To interpret the solution we make use of the conformal rescaling connecting 
the unit sphere to flat space, 

\begin{equation} ds^2_{\rm sphere} = \Omega^2ds^2_{\rm flat} \ , \end{equation}

\noindent where 

\begin{equation} \Omega^2 = \left( \frac{2r}{1+r^2} \right)^2 = (1+U)^2 \ . 
\end{equation}

\noindent If $\omega$ solves the conformal Laplace equation on the round 3-sphere then 

\begin{equation} \omega_{\rm BL} = \Omega^{\frac{1}{2}}\omega = \sqrt{1+U}
\omega \end{equation}

\noindent solves it on flat space \cite{Wald}. Thus the physical metric is given by 

\begin{equation} ds^2 = \omega^4ds^2_{\rm sphere}= \omega_{\rm BL}^4ds^2_{\rm flat} 
\ . \end{equation}

\noindent For the next interpretative step we adjust the coordinates so that the 
south pole (at $U = -1$) is placed at one 
of the punctures.  Thus the fourth component 
of the vector ${\bf X}_1$ equals $1$. Let the fourth components of the vectors 
${\bf X}_i$ be some fixed numbers $c_i$. 
Then we can use stereographic coordinates to calculate 

\begin{eqnarray} \omega_{\rm BL} = \sqrt{1+U}\omega = 1 + \sum_{i=2}^N
\sqrt{\frac{1+U}{1+X\cdot {\bf X}_i}} = \hspace{15mm} \nonumber \\ \\ 
= 1 + \sum_{i=2}^N\sqrt{\frac{2}{(1+r^2)(1+X\cdot {\bf X}_i)}} \sim 
1 + \sum_{i=2}^N\sqrt{\frac{2}{1-c_i}} \frac{1}{r} \ . \nonumber \end{eqnarray}

\noindent In the last step we assumed $r$ to be large, and then glanced at eqs. 
(\ref{embcoord}). We see that the puncture on 
the 3-sphere corresponds to an asymptotically flat end of the solution. 

We can also read off the Arnowitt--Deser--Misner mass. Using the standard definition 
\cite{Wald} we see that it equals twice the coefficient in front of the $1/r$ term. 
Relaxing the coordinate system it is 

\begin{equation} M_{\rm ADM} = 2\sum_{i=2}^N \sqrt{\frac{2}{1-{\bf X}_1\cdot {\bf X}_i}} 
= m\sum_{i=2}^N \sqrt{\frac{2}{1-{\bf X}_1\cdot {\bf X}_i}} \ . \label{ADM} \end{equation}

\noindent In the last step we momentarily reintroduced the scale factor from eq. 
(\ref{skala}). 

A more careful scrutiny is needed in order to establish that, when seen from the 
inside, the solution describes $N$ black holes each surrounded by a minimal surface. 
In fact this is not true for $N = 2$, in which case we have simply obtained the 
Schwarzschild solution in isotropic coordinates. Then there is no `inside', and 
only one minimal surface. For $N > 2$ the answer depends on how the punctures 
are distributed on the 3-sphere \cite{Brill}. For most of the solutions that we will 
consider we will be able to answer the question by inspection. 

Later on we will be interested in the local curvature of the solution. This 
is most easily worked out using embedding coordinates and the formulas in Appendix 
E of Wald \cite{Wald}. Full calculational detail is given elsewhere \cite{lic}. One 
finds that the Ricci tensor is 

\begin{eqnarray} R_{ac} = \frac{3}{2\omega^2}\sum_{i,j}\left[ \omega_i^3\omega_j^3
({\bf X}_{ia}+X_a)({\bf X}_{jc}+X_c) - \omega_i\omega_j^5({\bf X}_{ja}+X_a)({\bf X}_{jc}+X_c) 
+ \right. \nonumber \\ \\ 
+ \left. 
\frac{1}{3}\hat{g}_{ac}\omega_i^3\omega_j^3(1-{\bf X}_i\cdot {\bf X}_j) \right] \ . 
\hspace{35mm} \nonumber \end{eqnarray}

\noindent We will be especially concerned with two dimensional cross-sections 
of the initial data space. We define them by taking an equatorial slice of the 
3-sphere (such as $U = 0$). The induced metric is then 

\begin{equation} d\gamma^2 = \omega^4(d\theta^2 + \sin^2{\phi}) \ , \hspace{8mm} 
\omega = \sum_{i=1}^N\omega_i \ , \hspace{5mm} \omega_i = 
\frac{1}{\sqrt{1+X\cdot{\bf X}_i}} \ . \label{2metric} \end{equation}

\noindent This time $X^a$ and ${\bf X}^a_i$ are three dimensional vectors, 
the latter being constant and obeying 

\begin{equation} {\bf X}_i\cdot {\bf X}_i \leq 1 \ . \end{equation}

\noindent If we let $\Delta \omega$ and $(\nabla \omega )^2$ stand for the Laplacian 
and the gradient squared on the unit 2-sphere we find for the curvature of the 
induced metric that 

\begin{eqnarray} ^{(2)}R = \frac{1}{\omega^4}\left[ ^{(2)}\hat R - 
\frac{4\Delta \omega}{\omega} 
+ \frac{4(\nabla \omega)^2}{\omega^2}  \right] = \hspace{30mm} \nonumber \\ \nonumber \\ 
= \frac{1}{\omega^6}\left[ 3\omega\sum_i(1-{\bf X}^2_i)\omega_i^5 
- \sum_{i,j}\omega^3_i\omega^3_j(1-{\bf X}_i\cdot {\bf X}_j) \right] \ . 
\label{R} \end{eqnarray}

\noindent We observe that the result is a sum of two terms, each of which 
has a sign. The first is always positive, the second always negative. If all the 
punctures are placed on this 2-sphere then only the second term contributes, so the 
curvature of the cross-section is everywhere negative. If there are no punctures 
on the 2-sphere the curvature is positive on average. We study this in some detail in section \ref{local}.

\section{Black hole lattices}
\label{Platonic}
\noindent We must now decide how to place the punctures on the 3-sphere. 
Regularly, or at random? 

We will do both, and begin regularly \cite{Kjell, BandK, Wheeler}. One can then 
make use of a tessellation of the 3-sphere into identical cells, or equivalently 
place the punctures at the vertices of a four-dimensional Platonic body inscribed 
in the 3-sphere. The six Platonic bodies are well described in the literature 
\cite{Coxeter, Waegell}. The first example is the simplex. It was studied in detail 
(in this context) by Wheeler \cite{Wheeler83}, but since it has only $N = 5$ 
vertices we ignore the simplex here. Then comes the orthoplex, the four dimensional 
analogue of the octahedron. It has $N = 8$ vertices, conveniently placed 
at $\pm (1,0,0,0)$ and all their permutations. Alternatively, the vertices can 
be placed at $\pm (1,1,1,1)$ and all the permutations of $(1,1,-1,-1)$, or at 
$\pm (1,1,1,-1)$ and all their permutations. The orthoplex has altogether 16 facets 
or cells, and is therefore also known as the 16-cell. It is dual to the 
four-dimensional cube, also known as the tesseract. The latter has 8 cells, 
and can be obtained as the convex hull of two orthoplexes. The convex hull of all 
three of the listed sets of vertices (assumed normalized) is a self-dual Platonic 
body known as the icositetrachoron or 24-cell. It has no three dimensional analogue. 
Analogues of the icosahedron and the docecadron do exist in four dimensions. One of 
them has $N = 120$ vertices, including the vertices of the 24-cell as given above 
and 96 additional vertices placed at $(\pm \tau, \pm 1, \pm \tau^{-1}, 0)$ and all 
their even permutations, where $\tau$ is the Golden Mean. Having 600 cells it is 
known as the 600-cell. It is dual to the 120-cell which has $N = 600$ vertices. 
Conveniently the 120-cell also contains 10 copies of the 600-cell as subpolytopes. 
The list of Platonic configurations with $N$ black holes is thereby complete.  

Since we want to preserve the symmetries of the polytopes it now becomes evident 
why we were setting all the integration constants equal in eq. (\ref{sol}). The ADM 
mass measured in an asymptotically flat end corresponding to a puncture of the 
sphere is easily computed from eq. (\ref{ADM}). Specifically for the Platonic 
configurations we find 

\begin{equation} M_{\rm ADM} = \left\{ 
\begin{array}{ccl} 18.970\frac{m}{2} & \mbox{for} & N = 8 \\ 44.208\frac{m}{2} & \mbox{for} & N = 16 \\
69.445\frac{m}{2} & \mbox{for} & N = 24 \\ 386.438\frac{m}{2} & \mbox{for} & N = 120 
\\ 1985.19\frac{m}{2} & \mbox{for} & N = 600 \ . \end{array} 
\right. \label{ADMP} \end{equation}

%
%
%

\noindent We give the approximate numbers because they are the more illuminating. 
In our calculations we have set the dimensionful parameter $m = 2$.  

At the other end of the spectrum, starting at complete regularity, is complete 
randomness. We then begin with a definite vector $(1,0,0,0)$, say, and apply 
$N$ four dimensional rotation matrices chosen at random according to the Haar 
measure on the rotation group. A simple procedure for how to implement this 
is available \cite{Muller, Diaconis}. We keep all coefficients in eq. (\ref{sol}) equal, but because the symmetry is 
broken the ADM masses will now differ between the $N$ asymptotic regions. 
The average ADM mass measured at an asymptotic end of a random configuration, 
again averaged over 10 000 random configurations, is 

\begin{equation} \langle M_{\rm ADM} \rangle = \left\{ 
\begin{array}{ccl} 23.8\frac{m}{2} & \mbox{for} & N = 8 \\ 50.9\frac{m}{2} & \mbox{for} & N = 16 \\
78.1\frac{m}{2} & \mbox{for} & N = 24 \\ 404.0 \frac{m}{2} & \mbox{for} & N = 120 
\\ 2033.8\frac{m}{2} & \mbox{for} & N = 600 \ . \end{array} 
\right. \end{equation}

\noindent This is somewhat higher than for the regular configurations. 

One may wonder about the effects of clustering in the random configurations. The 
masses assigned to the individual black holes vary between roughly 1900 and 2250 
(times $m/2$) if $N = 600$, for the 10 000 examples we studied. The ones with 
the highest masses tend to have other black holes nearby. We did divide the sphere 
into 24 equal cells, and looked for a correlation between the number of black holes 
in a cell versus their masses averaged over the cell. The correlation exists, but is 
not very striking. Clifton \cite{Clifton} has studied clusters created using the method of 
images, but no direct comparison is possible because the latter method specifies 
the `bare masses', that is the integrations constants in eq. (\ref{sol}), in 
a different way. From our point of view it would be preferable to introduce clustering with statistical methods, as done in a recent paper by Jolin and Rosquist \cite{jr}. 

Another new feature is that if some punctures land too close to each other, then 
the solution contains fewer than $N$ black holes \cite{Brill}. Moreover, counting 
the number of minimal surfaces can become a complicated affair. The case of 
$N = 3$ punctures was studied in detail by Bishop \cite{Bishop}. For large $N$ it 
is clear from Korzy\'nski's theorem \cite{Korzynski}, as well as from the results 
reported in section 5 below, that this will happen only rarely. Anyway this does 
not affect the calculation of the ADM masses in the asymptotic regions. 
 
It remains to normalize our solutions in some reasonable way, in order to compare 
them with a $k = 1$ Friedmann dust universe at the moment of maximal expansion. 
For the Platonic configurations we can compare the length of a suitable curve at 
the boundary of a cell surrounding a black hole to that of a similar curve in the 
round Friedmann sphere \cite{Kjell, BandK}. This 
idea is not available for randomly chosen configurations, so instead we adopt 
a pragmatic suggestion due to Korzy\'nski \cite{Korzynski}. Averaging the conformal 
factor $\omega$ over the round 3-sphere we find (using suitable stereographic 
coordinates at the end) that 

\begin{equation} \langle \omega \rangle = \frac{1}{{\rm Vol(sphere)}} \int_{\rm sphere}\omega 
\ {\rm d}V = \frac{N}{2\pi^2}4\pi \int_0^\infty (1+U)^{\frac{5}{2}}r^2{\rm d}r = 
\frac{8\sqrt{2}N}{3\pi} \ . \end{equation}

\noindent We now choose the dimensionful parameter $m$ in the metric (\ref{skala}) 
so that 

\begin{equation} \frac{m^2}{4}\langle \omega \rangle^4 = 1 \hspace{5mm} 
\Leftrightarrow \hspace{5mm} m = \frac{9\pi^2}{64 N^2} \ . \end{equation}

\noindent Given that the physical metric $g$ approximates the round metric closely 
over most of the sphere, this is a reasonable way to choose a scale. If we removed 
the small regions around the black holes before performing the average then 
$\langle \omega \rangle$ would shrink somewhat, and $m$ would grow. But the 
calculation would no longer be easy to perform.

We can now compare the black hole lattices to a $k=1$ Friedmann universe at 
maximum expansion. There is only one sensible candidate for its total mass, namely  

\begin{equation} M_{\rm Fr} = V\rho = 2\pi^2a^3\rho \end{equation}

\noindent where $V$ is its volume and $\rho$ is the dust density. At 
maximum expansion this gives the metric 

\begin{equation} ds^2_{\rm Friedmann} = \frac{16M^2_{\rm Fr}}{9\pi^2}(d\chi^2 
+ \sin^2{\chi}(d\theta^2+ \sin^2{\theta}d\phi^2)) \ . \end{equation}

\noindent Since we have decided to work with unit spheres we set 

\begin{equation} M_{\rm Fr} = \frac{3\pi}{4} \ . \end{equation}

\noindent We can now ask how closely the sum of the ADM masses of the black hole 
lattice approximates $M_F$. If the ratio is close to unity this can be phrased 
as saying that the back reaction, in the sense of the averaging problem, is 
negligible \cite{Kjell, Korzynski}.

Of course it can be (and has been \cite{Clifton}) asked what the sum of the 
ADM masses as measured in the $N$ asymptotic regions has to do with the mass 
of the universe one finds inside. Actually there is an answer, at least for the 
Platonic configurations where---as we will see in section \ref{local}---the strong 
curvature regions surrounding the double sided marginally trapped surfaces are 
remarkably `round'. In other words, any irreducible mass associated with them 
would be very close to the ADM mass. A particular notion of quasi-local mass 
has been studied for these solutions, with reassuring results \cite{Tod}. 

For the Platonic configurations all the ADM masses are equal, and they were given in 
eqs. (\ref{ADMP}). Putting things together we find 

\begin{equation} \frac{N M_{\rm ADM}}{M_{\rm Fr}} = \left\{ \begin{array}{lll} 
0.698 & \mbox{if} & N = 8 \\ 0.813 & \mbox{if} & N = 16 \\ 0.852 & \mbox{if} & N = 24 \\
0.948 & \mbox{if} & N = 120 \\ 0.974 & \mbox{if} & N = 600 \ . \end{array} \right. 
\end{equation}

\noindent The inferred total masses are almost the same. In this sense the back reaction 
is indeed negligible \cite{Kjell}. For the random configurations 

\begin{equation} \frac{N \langle M_{\rm ADM}\rangle }{M_{\rm Fr}} = 
\left\{ \begin{array}{lll} 0.87 & \mbox{if} & N = 8 \\ 0.94 & \mbox{if} & N = 16 \\ 
0.96 & \mbox{if} & N = 24 \\0.991 & \mbox{if} & N = 120 \\ 0.998 & \mbox{if} & N = 600 
\ . \end{array} \right. \end{equation} 

\noindent Replacing the regular configurations with completely random ones has 
only a modest effect on these numbers. It just makes the agreement between the 
black hole lattice and the Friedmann universe a little bit better. 

\section{What does it look like?}
\label{lattice}

\noindent We now want to know what the black hole lattice spaces look like. For 
$N = 2$ they coincide with a $t = 0$ slice through the Schwarzschild solution, 
and the answer was given long ago in the form of an embedding diagram of a two 
dimensional slice known as Flamm's paraboloid. (Actually Flamm \cite{Flamm} drew 
only half of it, and left it to Einstein and Rosen \cite{Einstein} to discover the 
other half.) When $N > 2$ a similar picture is too much to hope for. What 
we can do, however, is to embed any space conformal to a sphere in a flat Minkowski 
space of two dimensions higher, as a cut of the lightcone which in standard coordinates 
is given by $T = R$, where $R$ is a radial Minkowski space coordinate. Thus we set 

\begin{equation} T = R = \omega^2 (\chi, \theta , \phi ) \ . \end{equation}

\noindent Then we project this embedding down to the $t = 0$ plane. Taking an 
equatorial slice of the projection results in rather interesting 
pictures with a clear meaning. They are somewhat hard to read however. The picture 
for $N = 2$ does not resemble Flamm's paraboloid except in superficial ways, 
while the picture for $N = 1$ ---which is actually a picture of flat space---is 
a paraboloidal surface of revolution. 

\begin{figure}[tbp]
\begin{center}
\vspace{-.5cm}
\includegraphics[scale=0.35]{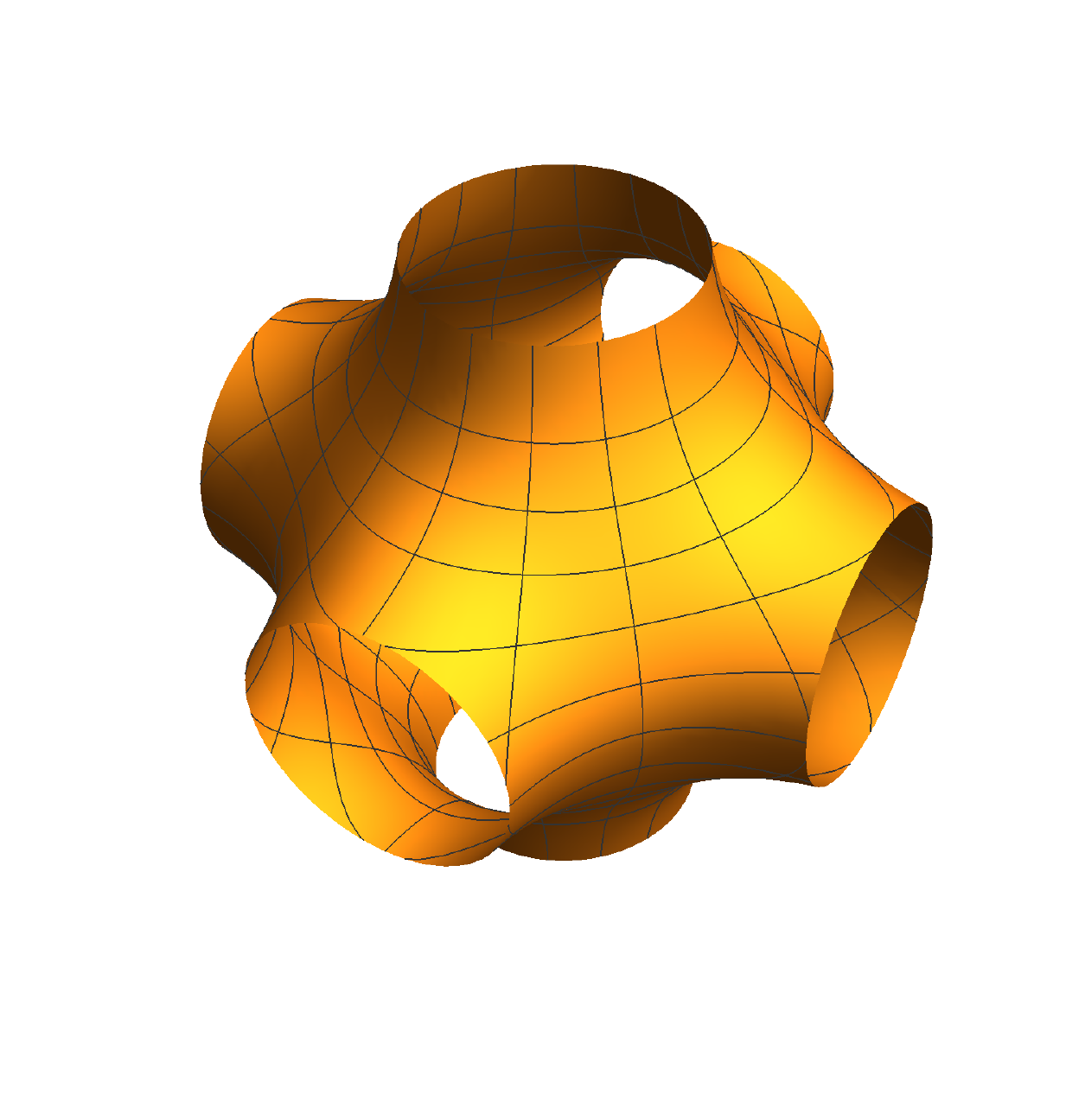}.\label{fig:a}\hfil
\includegraphics[scale=0.35]{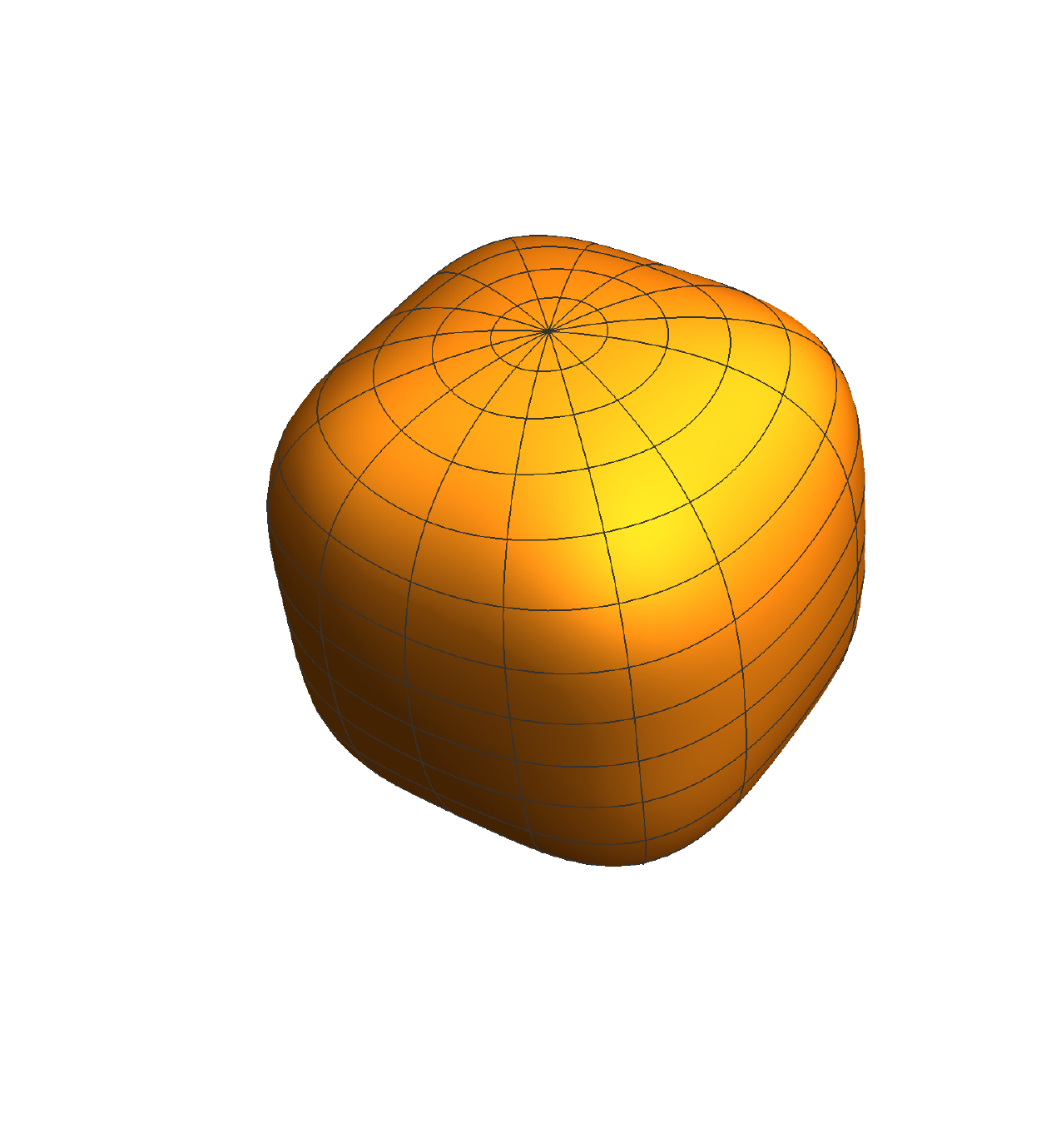}.\label{fig:b}
\vspace{-1cm}
\caption{\small{Two cross-sections of the orthoplex solution, embedded in a Minkowski 
space lightcone and then projected down to the $t=0$ hyperplane. Left: In this cross-section we see six out of eight black holes. 
Right: In this cross-section no black hole is seen, but we see six cell centres maximally distant from the black holes.}}
\label{6bh}
\end{center}
\end{figure}


With this reservation made, we present this picture for two different slices 
of the $N = 8$ Platonic solution in Figure \ref{6bh}. Further examples were displayed by Clifton 
et al. \cite{Kjell}, but in their plots 
they replaced $\omega^2$ with $\omega$ itself, making the pictures appear somewhat 
rounder at the price of losing their precise interpretation. The main point is the same 
though: when $N$ is large, say 120, the cross-sections reveal a body that looks like 
a round sphere except for sharp peaks in the immediate neighbourhood of the black holes. 
Thus, for a large portion of the 3-sphere, the physical metric $g$ is very close 
to the metric on a round 3-sphere. This observation was made fully precise, for more 
general configurations where $N$ can be chosen arbitrarily large, by Korzy\'nski 
\cite{Korzynski}. In the next section we will be concerned with 
the behaviour of the second derivatives of the metric, and the intrinsic curvature. 
This is very hard to read out from pictures of this nature. We will see that the intrinsic 
curvature is negative on the entire cross-section to the left in Figure \ref{6bh}.


\section{Local curvature}
\label{local}
\begin{figure}[bbp]
\begin{center}
\includegraphics[scale=0.5]{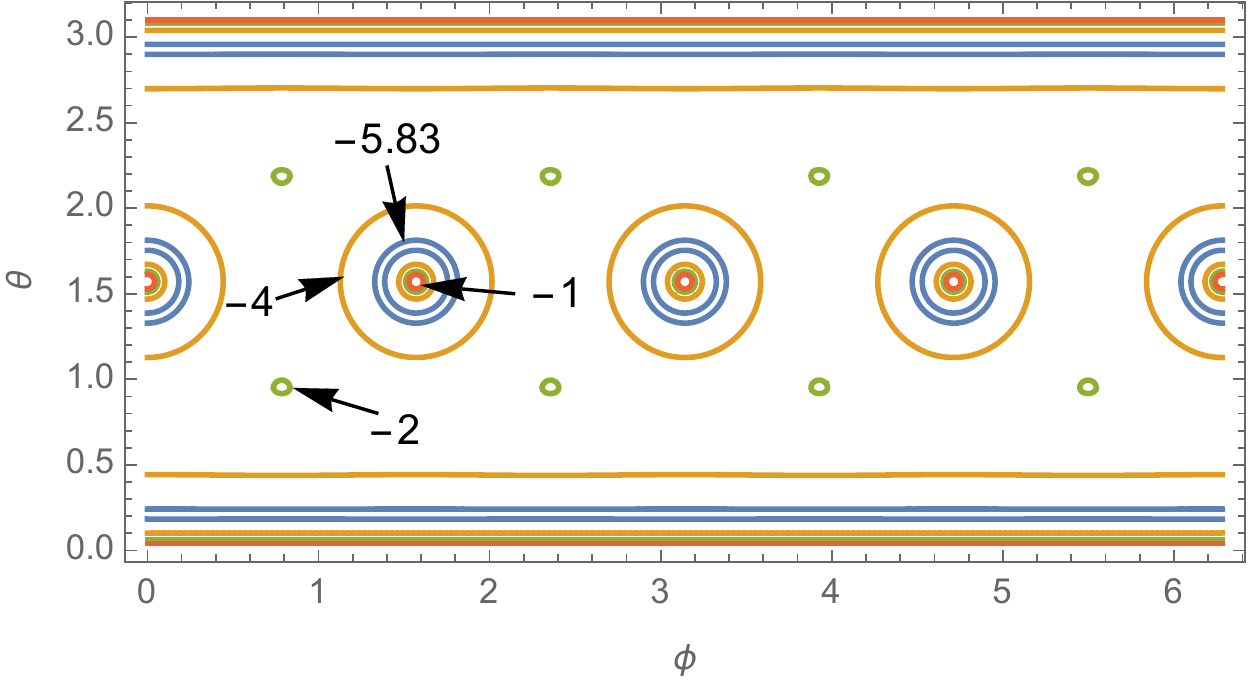}.\label{fig:a}\hfil
\includegraphics[scale=0.5]{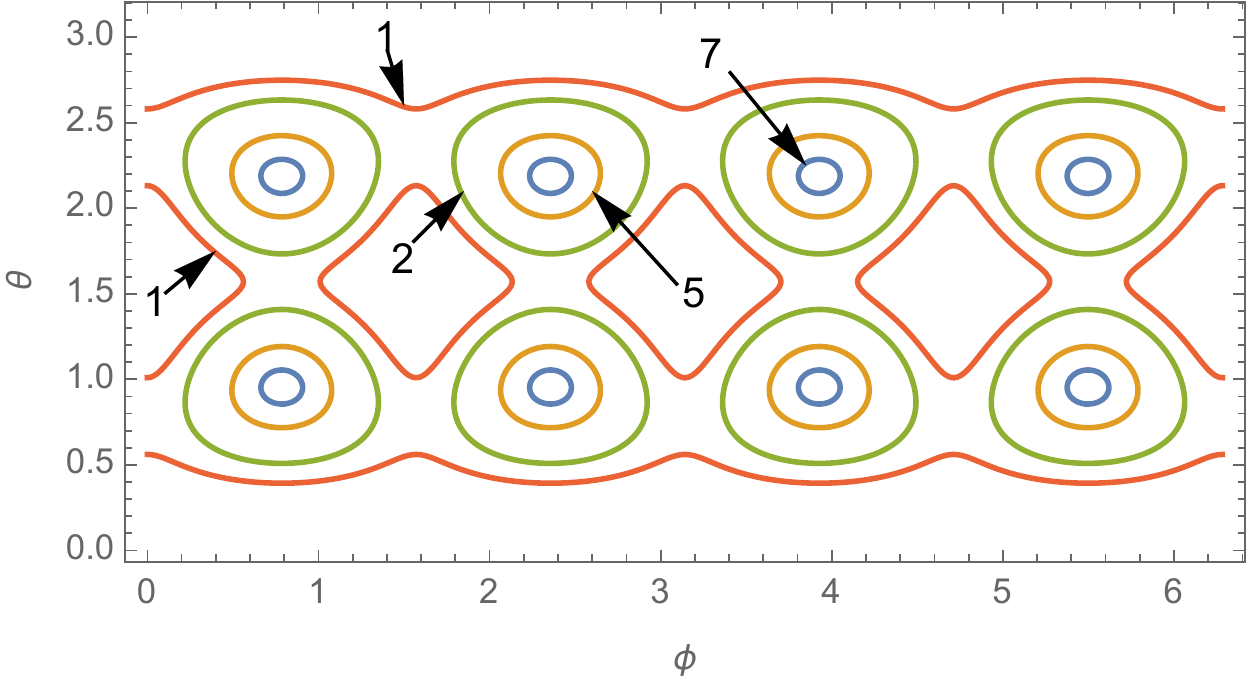}.\label{fig:b}
\vspace{-0.0cm}
\caption{\small{The Gaussian curvature on the two cross-sections shown in
Figure \ref{6bh}. The six black holes visible in the one on the left are remarkably round.
The curvature vanishes at the six cell centres seen (in corresponding positions)
on the right.}}
\label{contour8bh}
\vspace{-0.4cm}
\end{center}
\end{figure}

\noindent We stick to the idea of taking a two dimensional `equatorial' cross-section 
of our space, but this time we will look at Figure \ref{6bh} under the microscope provided by 
Gaussian curvature. The induced metric on the slice is given by eq. (\ref{2metric}), 
and its curvature scalar in eq. (\ref{R}). The coordinate system is adapted so that 
$\theta$ and $\phi$ serve as coordinates on the two dimensional cross-section. At 
first we set $N = 8$. Figure \ref{contour8bh} shows the behaviour of the function $^{(2)}R(\theta , \phi )$, for the same two cross-sections appearing in Figure \ref{6bh}. The normalization is such that we are comparing 
the black hole lattice with a Friedmann universe of unit radius. The unit 2-sphere 
has $^{(2)}R = 2$, and we see immediately that the black hole lattice is not close 
to this, or any other, round sphere as far as its local curvature is concerned. 
Nor should we have expected this to hold: the fitting is concerned with the metric 
$g$, not with its second derivatives. 

There are two more remarks to make about Figure \ref{contour8bh}. The black hole regions are 
located where the curvature assumes its minimum value ($^{(2)}R \approx - 5.9$), 
and it strikes the eye that they are well isolated from each other and do not distort each other 
noticeably. In fact at a cursory glance they appear to be perfectly round. We will 
soon quantify this. Meanwhile, on any cross-section free of punctures the average 
curvature must be positive for topological reasons. The reader may wonder how 
this manifests itself if the cross-section including the punctures is rotated just 
slightly so that the punctures disappear. The answer is that the curvature will 
become positive where the punctures used to be, and since the volume is large there 
the average curvature behaves as expected

\begin{figure}[bbp]
\begin{minipage}[c]{0.7\linewidth}
\includegraphics[width=\linewidth]{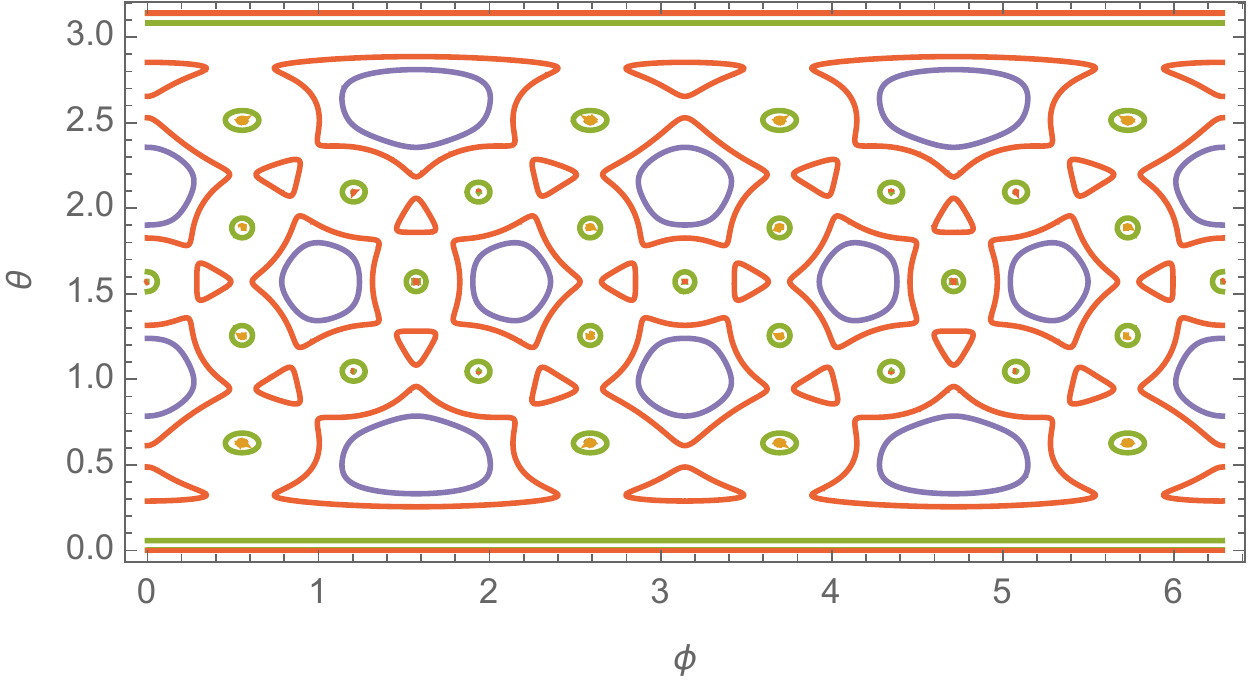}
\caption{\small{A slice through the configuration with 120 masses. There are
30 black holes residing inside the little green contours (at --100). The
red contour is at --2, and the blue at 0.}}
  \label{120bh}
\end{minipage}
\hfill
\begin{minipage}[c]{0.28\linewidth}
\includegraphics[width=\linewidth]{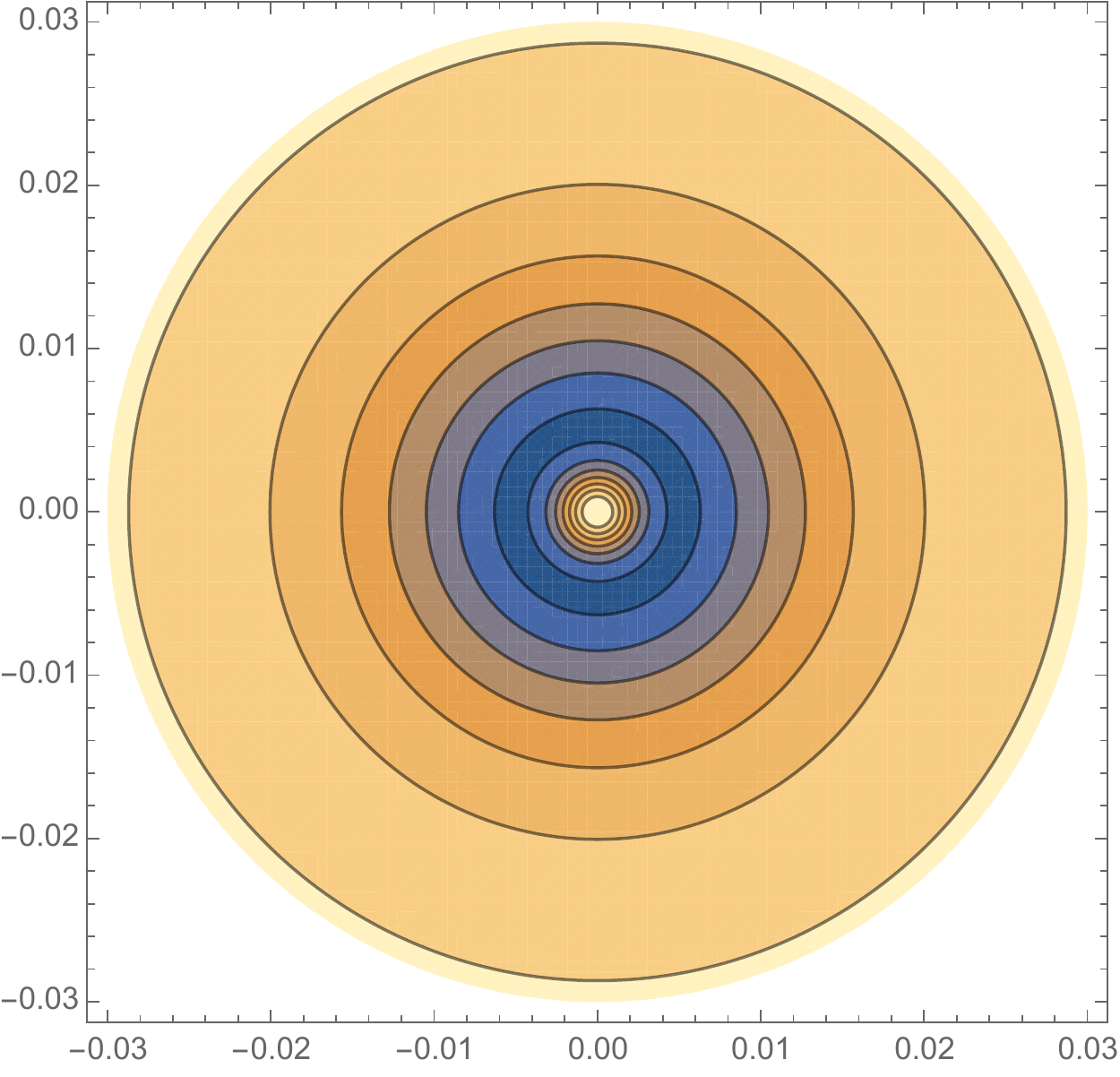}
 \caption{\small{A stereographic zoom-in (bounded by $^{(2)}R=-100$) on one of the black hole regions in Figure \ref{120bh}.}}
  \label{zoom}
\end{minipage}%
\end{figure}

The same conclusions hold for the larger Platonic configurations. In Figure \ref{120bh} we 
show a similar plot of a cross-section through the $N = 120$ lattice. The cross-section 
is again `equatorial', and the equator has been chosen in such a way that a pair of 
antipodally placed black holes reside at the poles of the 3-sphere. This means that 
30 black holes reside on the equator, and are visible in the picture \cite{Coxeter}. 
The curvature would have been hard to infer from a picture drawn according to the same recipe that produced
Figure \ref{6bh}. The latter seems to depict a round sphere with 30 sharp spikes protruding from it \cite{Kjell}. In Figure \ref{120bh} we do observe 12 isolated regions with
positive curvature. The curvature gradients in the neighbourhood of the
black holes is now so much larger than the gradients in the surrounding `universe' that we resort to a separate enlarged plot of a black hole region, using
stereographic coordinates for the purpose in order to exhibit the `roundness'
of the black hole shown. See Figure \ref{zoom}.

\begin{figure}[tbp]
\begin{center}
\vspace{-.5cm}
\includegraphics[scale=0.6]{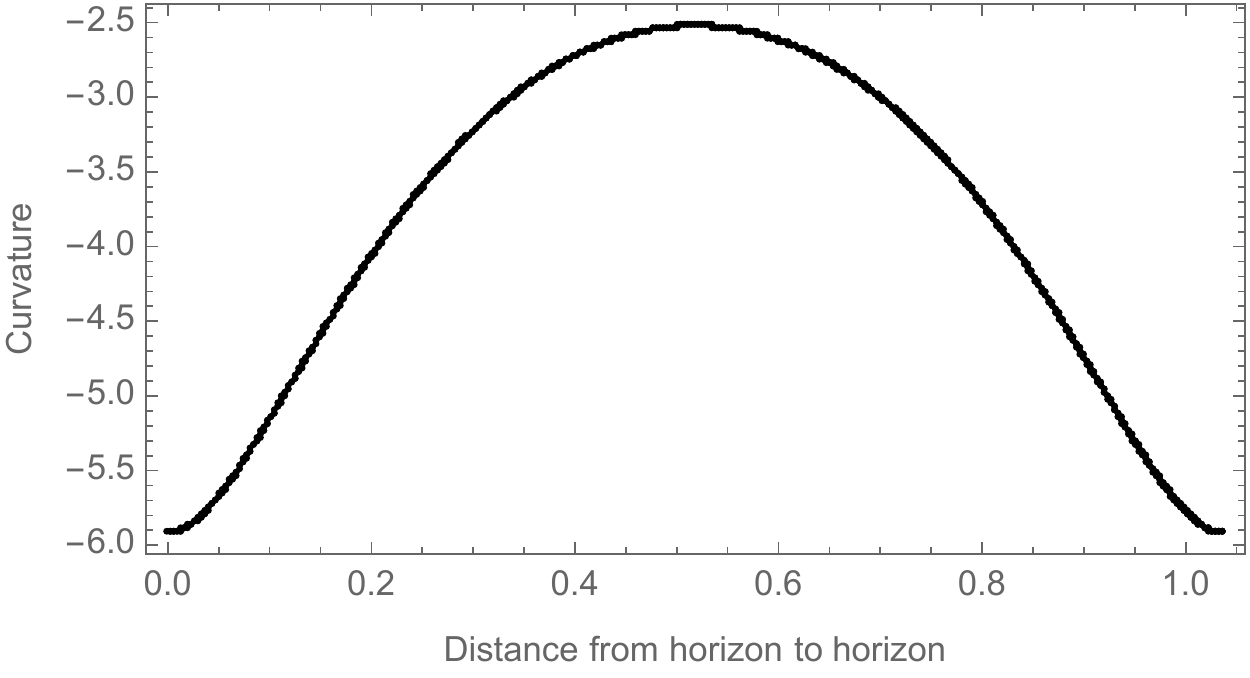}.\label{fig:a}\hfil
\includegraphics[scale=0.6]{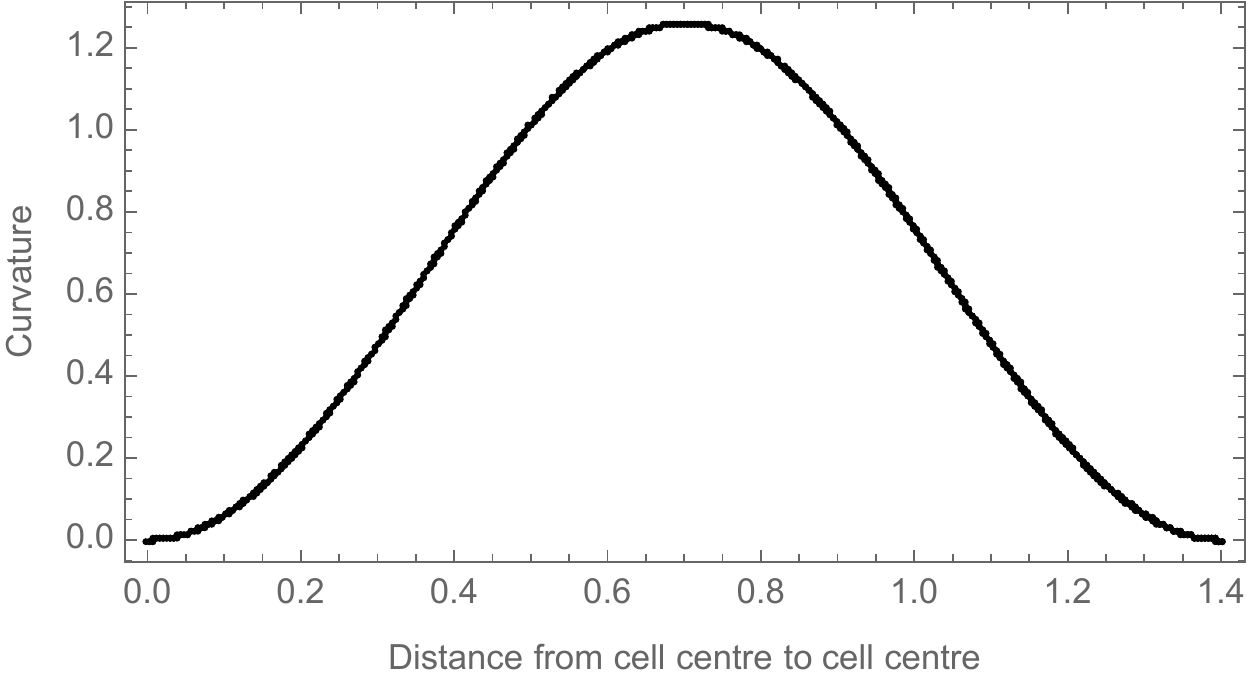}.\label{fig:b}
\caption{\small{The curvature scalar $^{(2)}R$ as a function of geodetic distance, 
for $N = 8$. Left: Along the shortest curve from one black hole horizon to another. 
The curve extends from one minimum of $^{(2)}R$ to another. Right: Along the shortest 
geodesic going from one cell centre to another. The curvature is zero at the 
cell centres and the total distance is close to $\pi/2$, as it would be on the unit 
sphere.}}
\label{BL3}
\end{center}
\end{figure}

So far our illustrations have been coordinate dependent. In Figure \ref{BL3} we show how 
the curvature varies with true distance along two geodesics in the Platonic $N = 8$  
configuration. One geodesic connects two black holes, the other two cell centres. 
Had we continued the geodesic into the asymptotic region it would not have been 
possible to see any difference between this part of the diagram and the same 
diagram drawn for the Schwarzschild solution. There are no surprises there.

To put a precise number on the degree to which the black holes approximate the Schwarzschild geometry we observe that each asymptotic region
is ‘surrounded’ by a sphere on which the squared Ricci tensor assumes a
maximum when the sphere is crossed in a transversal direction. For the Schwarzschild black hole there holds

\begin{equation} \frac{32}{3}M_{\rm ADM}^4(R_{ab}R^{ab})_{\rm max} = 1 
\ . \end{equation}

\noindent This observation is scale invariant, and does not depend on the absolute 
value of the mass. In a black hole lattice there will be distortions, causing this 
quantity to depend a little on where on the sphere it is evaluated. For the 
Platonic $N = 8$ configuration we 
find 

\begin{equation} 
0.99940 \leq \frac{32}{3}M_{\rm ADM}^4(R_{ab}R^{ab})_{\rm max} \leq 1.00235 \ .
\end{equation}

\noindent The upper bound is reached at a point minimally distant from another black hole 
in the cross-section shown above. The lower bound is actually based on an informed guess 
and we cannot guarantee the last decimal. Both of these numbers are very close to $1$, 
which gives a quantitative way of saying that the geometry around a puncture is very 
close to that of a Schwarzschild black hole. Bentivegna and Korzy\'nski \cite{BandK} 
have already estimated the `roundness' of the minimal surfaces 
surrounding the black holes, so we have simply confirmed in a different way 
that in the regular lattices the black holes are indeed very round already for 
$N = 8$. The approximation gets even better as $N$ increases. 

\begin{figure}[tbp]
\centering
\begin{subfigure}{.5\textwidth}
  \centering
  \includegraphics[width=.9\linewidth]{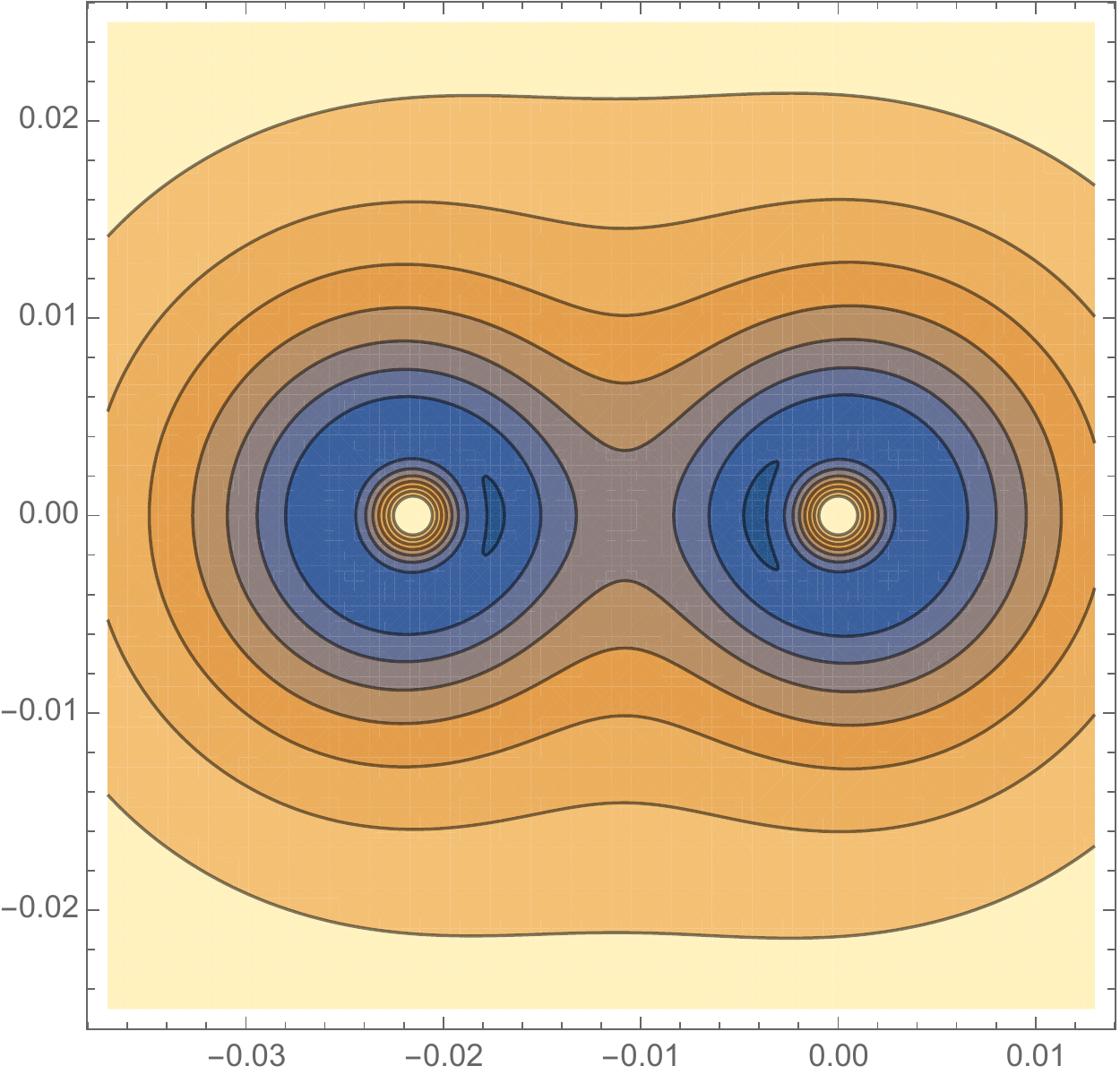}
  \caption{The ADM masses are 488 and 490.}
  \label{fig:sub1}
\end{subfigure}%
\begin{subfigure}{.5\textwidth}
  \centering
  \includegraphics[width=.9\linewidth]{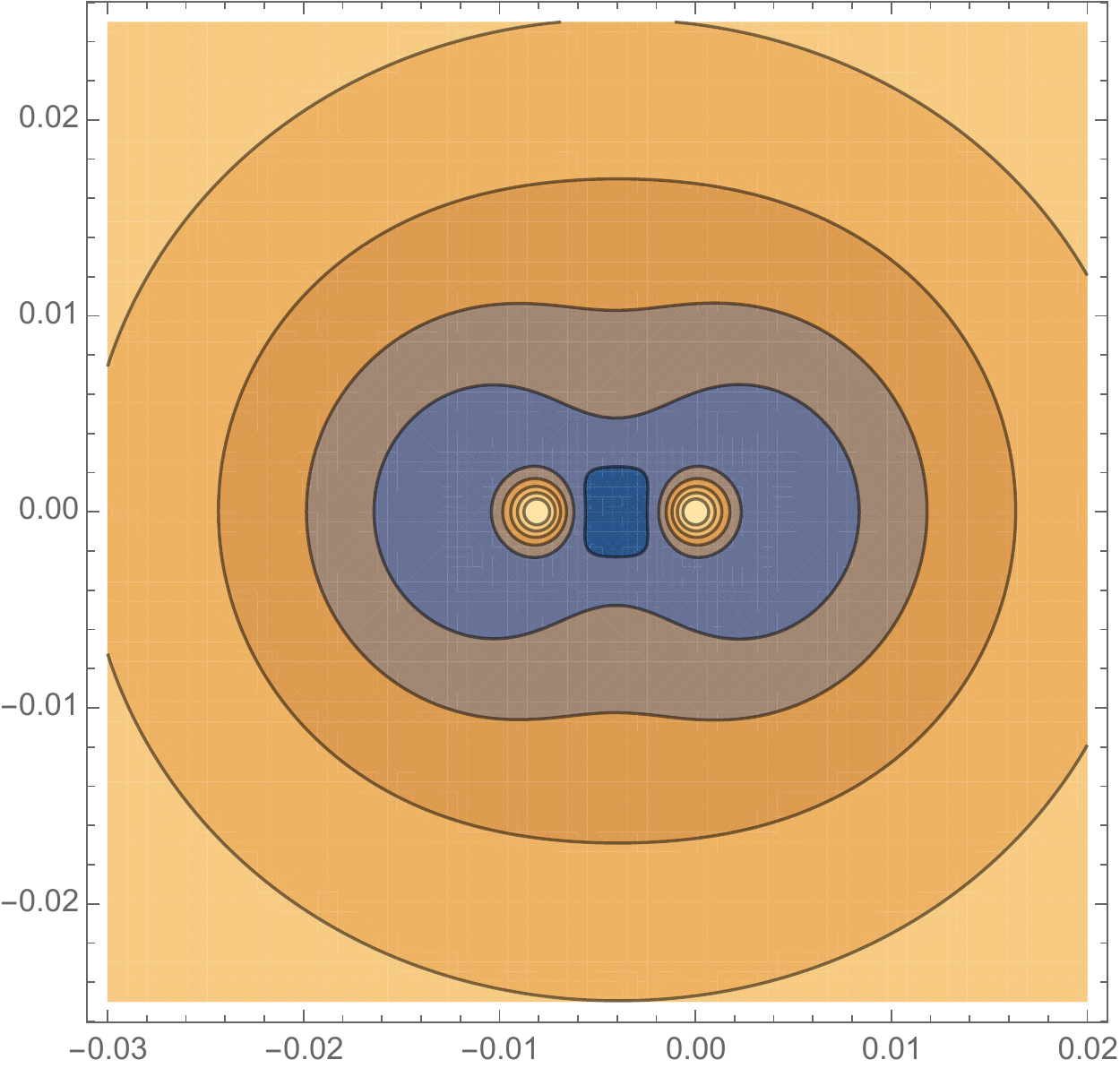}
  \caption{Both ADM masses are 636.}
  \label{fig:sub2}
\end{subfigure}
\caption{\small{Two examples of unusually close nearest neighbours in $N =120$ random configurations. Stereographic coordinates are used, and the ADM masses are given in dimensionless units. Compare to the well isolated black hole in Figure \ref{zoom}.}}
\label{random}
\end{figure}

For the random configurations the situation is of course harder to summarize. 
Evidently it can happen that two punctures land so close to each other that 
the black holes are significantly distorted, and indeed a puncture can 
lose its surrounding minimal surface \cite{Brill, Gibbons, Bishop}. The 
question is whether this is a rare phenomenon, or not. We looked briefly into this 
for the case $N = 120$. One look at how the local curvature behaves is enough to 
convince us that for the Platonic lattice each individual punture is surrounded 
by a spherical trough of negative curvature. See Figure \ref{zoom}. The local geometry on this two
dimensional cross-section is extremely close to that of Flamm's paraboloid 
once we have passed the trough and entered the asymptotic region. In fact we find, 
for the minimum of $^{(2)}R$, that 
\begin{equation} -4M^2_{\rm ADM}{^{(2)}R}_{\rm min} \approx 1.0000000000 \ , 
\end{equation}

\noindent where the eleventh decimal depends on how the cross-section is chosen. 
For Schwarzs--child the value is $1$. We then picked the position of the 120 
punctures at random, and among them we picked the three punctures closest to 
each other on the 3-sphere in order to witness the maximal amount of distortion. 
The three vectors 
span a three dimensional subspace of ${\bf R}^4$, and hence define a unique 
equatorial cross-section of the 3-sphere passing through these three punctures. 
Figure \ref{random} shows two examples where the nearest neighbours are unusually close, 
and their ADM masses are unusually large. (Recall that $M_{\rm ADM} \approx 
386$ for the Platonic configuration.) In example \ref{fig:sub1}, each 
black hole is surrounded by its own spherical trough of strongly negative 
curvature. To see how close it is to Schwarzschild in this region we evaluated 
the minimum curvature, and found that 
\begin{equation} -4M^2_{\rm ADM}{^{(2)}R}_{\rm min} \approx 1.032 .  
\end{equation}

\noindent The maximal value is reached inside one of the little lunes that are 
visible between the black holes. In example \ref{fig:sub2} there is only a 
single curvature minimum separating the two asymptotic regions. We looked through 100 different random configurations in this way, and found only 3 examples of the latter kind of behaviour. Based on this we dare to say that 
already for $N = 120$ the `typical' black hole lattice consists of 120 well separated black holes. 

\section{Concluding remarks}
\label{concl}
\noindent We have reported that the main conclusions regarding regular Platonic black
hole configurations change very little if the configuration is chosen at random
(in the sense of the Haar measure on the rotation group). The cosmological back reaction, as defined by Clifton 
et al. \cite{Kjell}, is even smaller for random configurations. Our main purpose was 
to study the second derivatives of the metric as manifested in local curvature. As 
expected, the black hole lattices do not look like round spheres at this microscopic level. 
On the other hand we can confirm that the black holes themselves are remarkably round, 
and differ very little from spherically symmetric black holes in the strong 
curvature regions. This is reassuring. 

The recent surge of interest in these solutions was driven by the cosmological 
averaging problem. Is it likely to affect our understanding of the dark side 
of the universe in a significant way? The answer is disputed \cite{Green, Buchert}. 
We feel that the right way to go may well be to find other interesting toy models, 
where agreement can be reached quickly.  

\

\

\noindent \underline{Acknowledgements}: We thank Kjell Rosquist for attracting us 
to the subject. IB also thanks Miko{\l}aj  Korzy\'nski for some explanations.

\end{document}